\documentclass[aps,11pt]{revtex4-1}

\newtheorem{theorem}{Theorem}
\newtheorem{proposition}{Proposition}
\newtheorem{lemma}{Lemma}
\newtheorem{corollary}{Corollary}

\usepackage{graphicx}
\usepackage{amsfonts}
\usepackage{amsmath}
\usepackage{tabulary}
\usepackage{mathrsfs}
\usepackage{eucal}
\usepackage{appendix}
\usepackage[ruled]{algorithm2e}
\usepackage{natbib}

\newcommand\bR{\mathbb{R}}
\newcommand\bE{\mathbb{E}}

\newcommand\mO{\mathcal{O}}
\newcommand\mC{\mathcal{C}}
\newcommand\mG{\mathcal{G}}
\newcommand\mF{\mathcal{F}}

\newcommand\mP{\mathcal{P}}

\newcommand\mL{\mathcal{L}}

\newcommand\mD{\mathcal{D}}

\newcommand\bC{\mathbf{C}}

\def\o{\omega}

\newcommand\qed{$\square$}

\begin{document}

%Optimization of Additional Information Acquisition in Decision Making Problems: Main Framework

\title{Towards the full information chain theory: expected loss and information relevance}
\author{E. Perevalov}
\email[E-mail: ]{eup2@lehigh.edu}
\author{D. Grace}
\email[E-mail: ]{dpg3@lehigh.edu}
\affiliation{Lehigh University\\ Bethlehem, PA}
\date{\today}

\begin{abstract}
When additional information sources are available, an
important question for an agent solving a certain problem is how to optimally use the information the sources are capable of providing.  A  framework that relates information accuracy on the source side to information relevance on the problem side is proposed. An optimal information acquisition problem is formulated as that of question selection to maximize the loss reduction for the problem solved by the agent. A duality relationship between pseudoenergy (accuracy related) quantities on the source side and loss (relevance related) quantities  on the problem side is observed.
\end{abstract}



\maketitle

\section{\label{s:intro}Introduction}
When uncertainty is present, several approaches to decision making are used depending on the particular details of the problem being solved. If the main difficulty lies in a large number of possible solutions and a complex structure of the feasible region, optimization methods are usually used. If the number of possible solutions is relatively small and the main difficulty lies in the process of updating the initial information, decision theoretic methods are appropriate. In Markov decision processes and stochastic optimal control, additional assumptions (such as Markovian or Gaussian property) are made which allows one to obtain solutions with special properties making it possible to handle the dynamic aspect of the problem efficiently.

\begin{figure}
\includegraphics[scale=1.0]{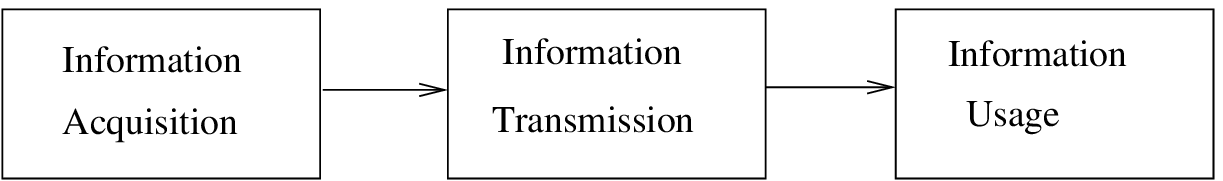}
\caption{\label{f:Ichain}The full information chain.}
\end{figure}

Regardless of the particular solution method, however, the common story of all such problem is information, or, more specifically, the lack thereof. In its current state, the fundamental quantitative theory of information is represented by Information Theory which, in spite of a number of fruitful connections with a variety of fields, is still predominantly a theory of information transmission. As such, it is concerned with information {\it quantity} and largely (if not entirely) oblivious to the possible content of information, including its {\it accuracy} and {\it relevance} to any kind of a problem. On the other hand, in many of its practical applications, the primary role of information lies in its ability to influence the quality of various decisions. It is clear that the ability of information to play this role depends critically not just on its quantity, but on its accuracy (with respect to describing the ``true state of affairs'')  and relevance (with respect to the particular problem). Put slightly differently, in its typical applications, information is acquired, then (possibly) transmitted and finally used to solve a certain problem. This typical path of information can be termed the {\it full information chain} (see Fig.~\ref{f:Ichain} for an illustration) which currently lacks\footnote{Part of the reason for such an omission is likely that the transmission link can be considered independently of the other two and that the nature (content) of the information does not play any role in solving the optimal transmission problem thus allowing for a universal and elegant treatment. On the contrary, it appears that the two ``end links'' of the information chain have to be optimized together and are very heterogeneous by nature thus making a universal assumption-free treatment problematic.} its basic fundamental theory with the sole exception of the middle (transmission) link.

This article is part of an effort to extend the classical Information Theory to a theory of the full information chain -- including the two ``end links''. Since these two links appear to be logically closely connected, the proposed extension has to take a form of a single joint theory meaning, in particular, that any design decisions (similar to source coding of classical Information Theory) can only be made when both links for the particular problem are taken into account. Still, due to the sheer volume of this task, it would appear reasonable to approach it in steps. Correspondingly, a quantitative description on the information acquisition link was addressed in \cite{part1,part2} where the process of information exchange between an agent (decision maker) and an information source was considered. This article's goal is to provide a similar treatment of the information usage link in a general setting. To make this task a bit more specific, the third link of the information chain is considered from the Operational Research perspective in that the main problem the agent is assumed to be solving is taken to have the form of a typical stochastic optimization problem with an objective in the form of an expected value.

\subsection{Related work}
The present work can be looked upon as an attempt to extend the classical Information Theory to make it useful for optimization and decision making under uncertainty. The field of Information Theory, born from Shannon's work on the theory of communications \cite{SHANNON:1948} since had great success in a number of fields. At the present time it would be impossible to attempt making any sort of comprehensive or even representative list of references pertaining to applications of Information Theory in communications and other fields. To mention a few more or less randomly selected examples, one could cite applications in statistical physics \cite{JAYNES:1957a,JAYNES:1957b}, computer vision \cite{viola1995}, climatology \cite{mokhov2006,verdes2005}, physiology \cite{katura2006} and neurophysiology \cite{chavez2003}. The relatively new field of Generalized Information Theory (see e.g. \cite{klir1996})
is concerned with problems of characterizing uncertainty in frameworks that are more general than
classical probability such as Dempster-Shafer theory \cite{shafer1976}. In particular, it was shown in \cite{maeda1993,harmanec1994} that the minimal uncertainty measure satisfying consistency requirements is obtained by maximizing Shannon entropy over all classical probability distributions consistent with the given (generalized) belief specification.

As was mentioned earlier, this article is part of an effort to extend the domain of Information Theory to include information acquisition and usage processes. The former of these was previously addressed in the classical work of Cox \cite{cox1946,cox1961,cox1979} on the foundations of probability and theory of inquiry. This line of work received further development recently resulting in a formulation of the calculus of inquiry \cite{knuth05,knuth08} that, in particular, constructs a distributive lattice of questions dual to the Boolean lattice of logical assertions. The definition of questions adapted in \cite{part1} corresponds to the particular subclass of questions -- the partition questions -- defined in \cite{knuth05}. Our work in \cite{part1,part2} goes beyond that on the calculus of inquiry in that it introduces the concept of {\it pseudoenergy} as a measure of source specific difficulty of various questions to the given information source. One could say that it develops a quantitative theory of {\it knowledge} as opposed to the theory of information.

Information Physics \cite{caticha-rev} is a relatively new branch of physical sciences that studies the role information plays in fundamental laws of nature. This line of research goes back to the defining work of Jaynes \cite{JAYNES:1957a,JAYNES:1957b} on the application of the Principle of Maximum Entropy (MaxEnt) to derive the fundamental laws of thermodynamics. It is related to the proposed framework in that it addresses information {\it relevance} in application to physical sciences. The main Information Physics hypothesis is that the laws of nature are essentially the laws of inductive inference correctly applied to respective systems. In order to correctly formulate them one needs to know the degrees of freedom and the relevant information necessary to completely specify the system state. Recently, this approach (in modified and extended form) was applied to derive the fundamental laws of classical \cite{caticha07} and quantum \cite{caticha11} mechanics.

%(stochastic \cite{BIRGE:1997}, robust \cite{BENTALNEM:1998,BENTALBOYDNEM:2006} or, more recently, risk-averse \cite{eichhorn2005,ruszczynski2006b})

The idea of obtaining additional information to improve quality of decisions in situations characterized with uncertainty is obviously not new and has been pursued, for instance, in the area of statistical decision making. Applications to innovation adoption \cite{mccardle1985}, \cite{jensen1988}, fashion decisions \cite{fisher1996} and vaccine composition
decisions for flu immunization \cite{kornish2008} can be mentioned in this regard. Some
authors \cite{fischer1996}, \cite{ellison1993} introduced various models
(e.g. effective information model) for accounting for the actual, or effective, amount of
information  contained in the received observations. The common
theme of this line of work is in trying to find an optimal trade-off between the amount of additional
information obtained and the suitably measured degree of achieving the original goal. The difference of the proposed approach is in that it explicitly describes and allows to optimize over not
just the quantity of additional information but also its content and is based on explicit description of properties of information sources.

Explicit modeling of information sources that lies at the base of the proposed methodology is similar in spirit to analyzing and using information provided by human experts. In many practically relevant applications, the role of information sources will likely be played by human experts. In existing research literature, the problem of optimal usage of information obtained from experts has been addressed mostly in the form of updating the agent's beliefs given probability assessment from multiple experts \cite{french1985,genest1986,clemen1987,clemen1999} and optimal combining of expert opinions, including experts with incoherent and missing outputs \cite{predd2008}. In the framework developed in the present and related articles \cite{part1,part2}, the emphasis is on optimizing on the particular type of information for the given information source and a decision making problem.

\subsection{Outline}
In Section~\ref{s:1link}, we summarize the necessary information about the information acquisition link of the full information chain. Further details are given in Appendix~A. In Section~\ref{s:maps}, we describe maps from the parameter to the solution space and their properties that will be used later. Section~\ref{s:3link} contains the main part of the article -- a quantitative framework for the description of the information usage link of the full information chain. In Section~\ref{s:example}, we consider a simple example. Finally, Section~\ref{s:conclusion} contains a conclusion and a brief discussion of future developments. Appendix~B provides some of the longer proofs, and Appendix~C gives additional examples to illustrate some concepts introduced in the main text.

\section{\label{s:1link}Information Acquisition Link}
An agent is assumed to be interested in solving a problem. The latter is necessary to provide a context for information {\it relevance}. While the nature of the problem can in principle be arbitrary, it has to allow for a quantitative characterization of the solution quality, or, equivalently, loss (compared to a that achievable in the presence of complete information). To make the discussion a bit more specific, we take the problem to be of the following general form.
\begin{equation}
 \label{eq:gen_stoch}
 \mbox{min}_{x\in X} \int_{\Omega} f(\omega, x) P(d\omega).
 \end{equation}
 Here $X\subset \mD$ is the set of all feasible solutions, i.e. the set satisfying all (deterministic) constraints that are present in the problem formulation, where $\mD$ is the space to which all solutions belong (e.g. a suitable Euclidean space).  $\Omega$ has the meaning of a space of possible values of input data parameters that are not known with certainty. It is often referred to as a parameter space. $P$ is a fixed initial probability measure (with a suitable sigma-algebra $\mF$ assumed) on $(\Omega,\mF)$ that describes the initial state of information available to the agent. The function $f$: $\Omega\times \mD\rightarrow \overline \bR$ is assumed to be integrable on $\Omega$ for each $x\in X$. For example, in the context of stochastic optimization, $X$ is the set of feasible first-stage solutions and $f(\omega,x)$ is the best possible objective value for the first stage decision $x$ in case when the random outcome $\omega$ is observed.

The natural form of the loss for the formulation (\ref{eq:gen_stoch}) is
\begin{equation*}
L(P)=\int_{\Omega} f(\o,x^*_P) P(d\o)-\int_{\Omega} f(\o,x^*_{\o}) P(d\o),
\end{equation*}
where $x^*_P$ is a solution of (\ref{eq:gen_stoch}) and $x^*_{\o}$ is a solution of $\mbox{min}_{x\in X} f(\o,x)$ for the given $\o$. The agent's ultimate goal is in minimizing the loss given the available information source(s). To achieve that goal, the agent engages in information exchange with the source. This exchange constitutes the content of the first link of the information chain. In the course of the information exchange, the agent poses questions and the information source provides answers. The agent is assumed to be capable of ``deciphering'' the answers by mapping them to updated probability measures on $\Omega$. In Appendix~A, we present some details of the information exchange process. In particular, we briefly describe the notions of {\it question difficulty}, {\it answer depth} and {\it information source models} introduced in \cite{part1} and \cite{part2}.

\section{\label{s:maps}Maps and their properties}
In what follows, we make use of maps from $\Omega$ into $X$ with discrete image sets.
Let $\mG$ be the set of all such maps. Since the image set of all maps from $\mG$ is assumed to be discrete, any such map $g\in \mG$ can be uniquely described by the corresponding partition $\bC=\{C_1,\dotsc, C_r \}$ of $\Omega$ and the corresponding image set $I=\{x_1,\dotsc, x_r\}$ such that $g(\o)=x_j$ for all $\o\in C_j$. We will sometimes write $g=(\bC,I)$ whenever the components of a map (partition and image set) need to be made explicit.

The following maps from the set $\mG$ are important special cases that will be referred to later.
\begin{itemize}
\item Optimal (``zero loss'') map $g_0$:
$g_0(\omega)=x_{\omega}^*$, where $x_{\omega}^*$ is the solution of $\mbox{min}_{x\in X} f(\omega, x)$. It simply maps each scenario into the corresponding (deterministic) optimal solution.

\item All-to-one maps $g_x$: $g_x(\omega)=x$ for
all $\omega\in \Omega$. These map all elements of $\Omega$ into some single element of $X$.

\item For the given measure $P$ on $\Omega$, the stochastic optimal map $g_P$: $g_P(\omega)=x_P^*$, where $x_P^*$ is a solution of
(\ref{eq:gen_stoch}). Obviously, it is just a special case for of all-to-one maps $g_x$.

\item For the given measure $P$ and a (complete) partition $\bC=\{C_1,\dotsc, C_r\}$ of $\Omega$, the map $g_{P,\bC}$: $g_{P,\bC}(\o)=x_{P_{C_j}}^*$ for all $\o\in C_j$, $j=1,\dotsc, r$. (Here $x_{P_{C_j}}^*$ is an optimal solution of problem (\ref{eq:gen_stoch}) with measure $P$ replaced with the conditional measure $P_{C_j}$.) In the following, we denote by $\mC$ the set of all maps of the form $g_{P,\bC}$ for all possible partitions $\bC$ of $\Omega$ and will sometimes refer to maps from the set $\mC$ as {\it subset-optimal maps}.
\end{itemize}

Next, we define some useful functionals to be used later.

Let $P$ be any probability measure on $\Omega$ and $x$ an arbitrary element of the solution space $X$. We define the {\it suboptimality} of $x$ with respect to $P$ as follows:
\begin{equation}
\label{eq:subopt}
S(x,P)=\bE_P f(\omega, x)-\bE_P f(\omega, x_P^*)=\int_{\Omega} (f(\omega, x)-f(\omega, x_P^*)) P(d\omega),
\end{equation}
i.e. suboptimality of $x$ w.r.t. $P$ is the difference in objective values of problem (\ref{eq:gen_stoch}) if $x$ is used instead of the optimal solution $x_P^*$.

If $P$ is an arbitrary measure on $\Omega$ and $g\in \mG$ is an arbitrary map from $\Omega$ into $X$, we define the {\it loss} of $g$ with respect to $P$ as
\begin{equation}
\label{eq:loss}
L(g,P) = \bE_P f(\omega, g(\omega)) - \bE_P f(\omega, x_{\omega}^*)= \int_{\Omega} (f(\omega, g(\omega))-f(\omega, x_{\omega}^*)) P(d\omega).
\end{equation}
In particular, if $g=g_P$ is the stochastic optimal map corresponding to the measure $P$, the loss $L(g_P,P)$ is the traditional {\it expected value of perfect information} (EVPI). If $g=g_0$ is the optimal map, the loss is equal to zero for any measure $P$: $L(g_0,P)=0$.

Finally, for any  measure $P$ and map $g\in \mG$, we define the {\it gain} of $g$ with respect to $P$ as follows:
\begin{equation}
\label{eq:gain}
B(g,P)=\bE_P f(\omega, x_P^*)- \bE_P f(\omega, g(\omega)) = \int_{\Omega} (f(\omega, x_P^*)-f(\omega, g(\omega))) P(d\omega).
\end{equation}
The gain functional of a map $g$ measures the decrease in loss that can be achieved by the map $g$, compared to the best all-to-one map $g_P$. In particular, the largest possible gain obtains by an optimal map $g_0$, and for this map, the value of gain is equal to the loss of $g_P$, since any optimal map has zero loss. It is also clear that, while suboptimality and loss are always nonnegative, gain can take both positive and negative values. For example, the gain of any all-to-one map $g_x$ is negative unless $x=x_P^*$ (in which case the gain vanishes).

The following lemma states an elementary but useful relationship between gain and loss for an arbitrary map $g$ from $\Omega$ into $X$. The proof of the lemma is straightforward and therefore omitted.

\begin{lemma}
For any map $g\in \mG$ and any measure $P$ on $\Omega$,
$$B(g,P)+L(g,P)=L(g_P,P),$$
where $g_P$ is the stochastic optimal map for the measure $P$.
\label{l:B+L=L}
\end{lemma}

 The statement of Lemma~\ref{l:B+L=L} can be rewritten as $B(g,P)=L(g_P,P)-L(g,P)$ and, in fact can be used as a definition of the gain of arbitrary map $g\in \mG$: the gain is equal to the decrease of the value of loss compared to the loss of the best all-to-one map $g_P$.

 Let $f(\mP)\rightarrow \bR$ be a real-valued functional on the suitably restricted set $\mP$ of measures on $\Omega$. For the later developments it turns out to be convenient to introduce the following notation. Let $\bC=\{C_1,\dotsc, C_r\}$ be a partition of $\Omega$ (a question), and let $V(\bC)$ be an answer to $\bC$ that can takes values in the set $\{s_1, \dotsc, s_m\}$.

 We denote by $f(P_{\bC})$ the expected value of the functional $f(\cdot)$ over the set of conditional measures $\{P_{C_j}\}$, $j=1,\dotsc, r$:
\begin{equation}
f(P_{\bC})=\sum_{j=1}^r P(C_j)f(P_{C_j}),
\label{eq:f(bC)}
\end{equation}
and by $f(P_{V(\bC)})$ -- the expected value of $f(\bC)$ over the set of updated measures $\{P^k\}$, $k=1,\dotsc, m$:
\begin{equation}
f(P_{V(\bC)})=\sum_{k=1}^m \Pr(V(\bC)=s_k)f(P^k)=\sum_{k=1}^m v_k f(P^k),
\label{eq:f(V)}
\end{equation}

Then we can define suboptimality, loss and gain functionals for a given question $\bC$ and an answer $V(\bC)$ using the just introduced notational convention (\ref{eq:f(bC)}) and (\ref{eq:f(V)}).

Namely, for an arbitrary $x\in X$, the suboptimality of solution $x$ with respect to question $\bC$ (and initial measure $P$) is given by
\begin{equation}
S(x,P_{\bC})=\sum_{i=1}^s P(C_j) S(x,P_{C_j}),
\label{eq:S(bC)}
\end{equation}
and the suboptimality of $x$ with respect to answer $V(\bC)$ to question $\bC$ (and initial measure $P$) reads
\begin{equation}
S(x,P_{V(\bC)})=\sum_{k=1}^m v_k S(x,P^k).
\label{eq:S(V)}
\end{equation}

Likewise, for an arbitrary map $g\in \mG$, and question $\bC$, the loss and gain of $g$ with respect to $\bC$ are given by
\begin{equation}
L(g,P_{\bC})=\sum_{j=1}^r P(C_j) L(g,P_{C_j}),
\label{eq:L(bC)}
\end{equation}
and
\begin{equation}
B(g,P_{\bC})=\sum_{j=1}^r P(C_j) B(g,P_{C_j}),
\label{eq:B(bC)}
\end{equation}
respectively.

The loss and gain functionals for a  map $g\in \mG$ with respect to answer $V(\bC)$ are defined analogously:
\begin{equation}
L(g,P_{V(\bC)})=\sum_{k=1}^m v_k L(g,P^k),
\label{eq:L(V)}
\end{equation}
and
\begin{equation}
B(g,P_{V(\bC)})=\sum_{k=1}^m v_k B(g,P^k),
\label{eq:B(V)}
\end{equation}
respectively.

The following representation for the expected loss $L(g,P)$ will be useful later.
\begin{lemma}
For any map $g=(\bC,I)\in \mG$, the expected loss $L(g,P)$ can be written as
$$L(g,P)=\sum_{j=1}^r P(C_j)L(g,P_{C_j})=L(g,P_{\bC}).$$
\label{l:loss1}
\end{lemma}

{\bf Proof:} See Appendix~B. \qed

Let $g=(\bC,I)\in \mC$ be a subset-optimal map. Then the EVPI for the problem (\ref{eq:gen_stoch}) can be decomposed in a convenient way.

\begin{lemma}
For any map $g_{\bC,P}\in \mC$, the EVPI $L(g_P,P)$ of the problem (\ref{eq:gen_stoch}) can be decomposed as
$$L(g_P,P)=S(x_P^*,P_{\bC})+L(g_{\bC,P},P).$$
\label{l:L=S+L}
\end{lemma}

{\bf Proof:} See Appendix~B. \qed

\section{\label{s:3link}Information Usage Link}
In this section, a quantitative framework for the description of the third link of the full information chain is discussed. A connection to the first link is made resulting in a formulation of the optimal information acquisition problem.

\subsection{Pseudoenergy-loss efficient frontier}
Let us consider the set $\mG$ of maps from $\Omega$ into $X$. Each map $g=(\bC(g),I(g))$ from this set can be characterized by the corresponding loss $L(g,P)$ with respect to the original measure $P$ and the value $G(\Omega, \bC(g), P)$ -- the difficulty of the corresponding question. We will be interested -- for reasons that will become clear shortly -- in finding the {\it efficient frontier} in the Euclidean plane with coordinates $(G(\Omega, \bC(g), P), L(g,P))$. In other words, we will be looking for the set $\mO$ of Pareto-optimal maps that can be found by solving the following parametric optimization problem
\begin{equation}
\begin{aligned}
& \underset{g\in \mG}{\text{minimize}}
& & L(g,P)  \\
& \text{subject to}
& & G(\Omega, \bC(g),P)\le \gamma \\
\end{aligned}
\label{eq:Pareto}
\end{equation}
for all values of the parameter $\gamma$.

The first observation we can make is that to find the set $\mO$ of Pareto-optimal maps it is sufficient to consider the set of subset-optimal maps $\mC$ as the following proposition asserts.

\begin{proposition}
$\mO\subset \mC$
\label{p:OinC}
\end{proposition}

{\bf Proof:} Let $g=(\bC,I)$ where $I=\{x_1,x_2,\ldots, x_r\}$. Suppose that $g\notin \mC$. Then there exists at least one $C\in \bC$ such that $g(C)\ne x_{P_C}^*$. Without loss of generality we can assume that $C=C_1$. Consider a different map $g'=(\bC,I')$ such that $I'=\{x_{P_{C_1}}^*, x_2,\ldots, x_r\}$. Obviously, $G(\Omega,\bC(g'),P)=G(\Omega,\bC(g),P)$ (since $\bC(g')=\bC(g)$). On the other hand,
$$L(g',P)-L(g,P)=P(C_1)(L(g',P_{C_1})-L(g,P_{C_1}))<0, $$
since $L(g',P_{C_1})$ takes the minimum value among all maps with the same partition $\bC$. We thus find that $L(g',P)<L(g,P)$ which means that $g\notin \mO$. \qed

It follows from Proposition \ref{p:OinC} that one needs to look no further than the set $\mC$ of subset-optimal maps. Such maps are uniquely characterized by the corresponding partition $\bC$ only (up to simple equivalences). Therefore the task of finding maps that belong to the set $\mC$ is equivalent to that of finding the corresponding partitions of the set $\Omega$.

\subsection{Optimal information acquisition}
Let us now address the optimal information acquisition problem: what question(s) need to be asked the given information source in order to obtain the minimum possible loss for (\ref{eq:gen_stoch}). Given a question $\bC=\{C_1,\dotsc, C_r\}$ to an information source and its answer $V(\bC)$ taking values in the set $\{s_1,\dotsc, s_m\}$, we denote
by $\mL(s_k)$, $k=1,\dotsc, m$ the {\it minimum conditional expected loss} given that $V(\bC)=s_k$ and by  $\mL(V(\bC))$ the {\it minimum expected loss} that the agent can achieve given the answer $V(\bC)$. The latter can be found as
\begin{equation}
\mL(V(\bC))=\sum_{k=1}^m \Pr(V(\bC)=s_k)\mL(s_k),
\label{eq:mLV}
\end{equation}
i.e. as an expectation over possible values of the answer $V(\bC)$.

Clearly, if no answer was received -- and the agent has to choose a solution $x\in X$ based on the original information only -- the minimum expected loss will be equal to the EVPI of the original problem: $\mL(\emptyset)=L(g_P,P)$.

If the agent poses a question $\bC=\{C_1,\dotsc, C_r\}$ to the information source and receives a particular value $s_k$ of answer $V(\bC)$, the original measure $P$ on $\Omega$ gets updated to $P^k$. Therefore, in order to minimize loss for the given value $s_k$ of answer $V(\bC)$, the agent needs to choose the solution $x^*_{P^k}$ -- the solution minimizing the expectation $\bE_{P^k} f(\o,x)$ over all (feasible) values of $x$.

\subsubsection{Perfect answers}
First, let us assume that the information source can provide a perfect answer to $\bC$. Then the following result can be obtained.

\begin{proposition}
Let $\bC=\{C_1,\dotsc, C_r\}$ be a complete question and $g_{\bC,P}\in \mC$ be a corresponding subset-optimal map. If the agent is given a perfect answer $V^*(\bC)$ to $\bC$ then
$$\mL(V^*(\bC))=L(g_{\bC,P},P). $$
\label{p:loss-perf}
\end{proposition}

{\bf Proof:} See Appendix~B. \qed

Combining the result of Proposition~\ref{p:loss-perf} with Lemma~\ref{l:loss1} (valid for any $g\in \mG$)  and Lemma~\ref{l:L=S+L} (valid for any $g\in \mC$) we can find the value of the largest {\it loss reduction} due to a perfect answer to question $\bC$. The result is formulated as a corollary.

\begin{corollary}
Given a perfect answer to question $\bC$, the largest possible reduction in expected loss a agent can achieve is equal to
$$\mL(\emptyset)-\mL(V^*(\bC))=B(g_{\bC,P},P)=S(x^*_P,P_{\bC}), $$
where $g_{\bC,P}\in \mC$ is a subset-optimal map corresponding to question $\bC$.
\label{c:red-perf}
\end{corollary}

%%%%%%%%%%%%%%%%%%%%%%%%%%%%%%%%%%%%%%%%%%%%%%%%%%%%%%%%%%%%%%%%%%%%%%%%%%%%%%%%%%%

\subsubsection{Imperfect answers}
Now, let us relax the assumption of availability of a perfect answer to question $\bC$. Instead, we assume that the agent can obtain an answer $V(\bC)$ which is in general imperfect. First, we formulate a useful auxiliary result.

\begin{lemma}
Let $V(\bC)$ be an answer to question $\bC$ and let $g_{\bC,P}\in \mC$ be a corresponding subset-optimal map. Then
$$S(x_P^*,P_{\bC})=S(x_P^*,P_{V(\bC)})+B(g_{\bC,P},P_{V(\bC)}).$$
\label{l:S=S+G}
\end{lemma}

{\bf Proof:} See Appendix~B. \qed

Combining the result of Lemma~\ref{l:S=S+G} with that of Lemma~\ref{l:L=S+L}, we obtain a useful decomposition of the EVPI of the original problem which we formulate as a corollary.

\begin{corollary}
Let $V(\bC)$ be an answer to question $\bC$ and $g_{\bC,P}\in \mC$ a corresponding subset-optimal map. Then
$$L(g_P,P)=S(x_P^*,P_{V(\bC)})+B(g_{\bC,P},P_{V(\bC)})+L(g_{\bC,P},P). $$
\label{c:L=S+B+L}
\end{corollary}

Now we can determine the minimum expected loss $\mL(V(\bC))$ that's obtainable with the help of an answer $V(\bC)$ to question $\bC$. We state the result as a proposition.

\begin{proposition}
Let $\bC=\{C_1,\dotsc, C_r\}$ be a complete question and $g_{\bC,P}\in \mC$ be a corresponding subset-optimal map. If the agent is given a (generally imperfect) answer $V(\bC)$ to $\bC$ then
$$\mL(V(\bC))=B(g_{\bC,P},P_{V(\bC)})+L(g_{\bC,P},P). $$
\label{p:loss-imp}
\end{proposition}

{\bf Proof:} See Appendix~B. \qed

It is easy to see that, for perfect answer $V^*(\bC)$ to question $\bC$, the gain $B(g_{\bC,P},P_{V(\bC)})$ in Proposition~\ref{p:loss-imp} vanishes (since $B(g_{\bC,P},P_{V^*(\bC)})=B(g_{\bC,P},P_{\bC})=0$) and the result of Proposition~\ref{p:loss-perf} is recovered.

The amount of maximum reduction of loss due to answer $V(\bC)$ to question $\bC$ can be obtained by combining the result of Proposition~\ref{p:loss-imp} with that of Corollary~\ref{c:L=S+B+L}. The result is formulated as a corollary.

\begin{corollary}
Given a (generally imperfect) answer to question $\bC$, the largest possible reduction in expected loss a agent can achieve is equal to
$$\mL(\emptyset)-\mL(V(\bC))=S(x^*_P,P_{V(\bC)}). $$
\label{c:red-imp}
\end{corollary}

\subsection{Pseudoenergy-loss correspondence}
Comparing results obtained in this section with the corresponding pseudoenergy values discussed in Section~4 we can make several interesting observations regarding their correspondence that reveal a rather clear picture. We assume that the measure $P$ admits existence of a finest partition of $\Omega$. Let $C_f(P)$ be such finest partition. We can then summarize the observations made in the previous sections as follows.

\begin{itemize}
\item The initial loss is equal to EVPI $L(g_P,P)$. In order to reduce it to zero, one needs to completely resolve the underlying uncertainty by answering the exhaustive question $C_f(P)$ about possible outcomes on $\Omega$ perfectly. The required pseudoenergy is equal to $G(\Omega,\bC_f(P),P)$.

\item A perfect answer to question $\bC$ (that, as a partition, is some coarsening of $\bC_f(P)$) requires $G(\Omega,\bC,P)$ worth of pseudoenergy from an information source and allows the agent to reduce the loss by the amount equal to $S(x_P^*,P_{\bC})=B(g_{\bC,P},P)$.

\item If the source is able to produce only an imperfect answer $V(\bC)$ to question $\bC$ the corresponding amount of pseudoenergy is equal to the answer depth $Y(\Omega,\bC,P,V(\bC))$. Such an answer can reduce the initial loss $L(g_P,P)$ by the amount of $S(x_P^*,P_{V(\bC)})$.

\item The difference of depths (pseudoenergy contents) between a perfect and an imperfect answers to question $\bC$ is equal to $G(\Omega,\bC,P_{V(\bC)})$. The corresponding difference in loss reductions (values of information) is $B(g_{\bC,P},P_{V(\bC)})$. The latter quantity can be naturally interpreted as a price the agent pays for imperfection of the answer he/she receives to question $\bC$.

\item Given a perfect answer to question $\bC$, the residual pseudoenergy measuring the degree of difficulty of resolving the remaining uncertainty is equal to $G(\Omega,\bC_f(P)_{\bC},P)$. The corresponding residual loss is simply $L(g_{\bC,P},P)$.

\item Given an imperfect answer to question $\bC$, the residual pseudoenergy measuring the degree of difficulty of resolving the remaining uncertainty is equal to $G(\Omega,\bC_f(P),P_{V(\bC)})$ -- the difficulty of the exhaustive question $\bC_f(P)$ given the answer $V(\bC)$ to question $\bC$. The corresponding residual loss is equal to $\sum_{k=1}^m v_k L(g_{P^k},P^k)$.
\end{itemize}

Table~\ref{t:energy-loss} shows the correspondence between pseudoenergy and loss related quantities discussed above. We see that for every loss related quantity there is a corresponding pseudoenergy quantity, meaning that in order to reduce the loss by a certain amount the corresponding pseudoenergy has to be made available in the form of an answer to some question. Depending on the structure of the question, the amount of loss reduction and, respectively, the amount of residual loss can vary in size. The goal of the agent is to find the specific question(s) that would maximize the effect of the given information source (characterized by its pseudoenergy functional and source model parameters such as capacity) on the given problem. More specifically, the agent would want to find the specific question $\bC$ that would result in the smallest possible minimum expected loss $\mL(V(\bC))$ where $V(\bC)$ is the answer that the source can provide to question $\bC$. Formally, this {\it information acquisition optimization} problem can be written as
\begin{equation}
\begin{aligned}
& \underset{\bC}{\text{minimize}}
& & \mL(V(\bC))  \\
& \text{subject to}
& & Y(\Omega,\bC,P,V(\bC))=h(G(\Omega,\bC,P))\\
\end{aligned}
\label{eq:minLV}
\end{equation}
where minimization is performed over all possible partitions of the parameter space $\Omega$. The expression for the minimum loss $\mL(V(\bC))$ is given either by Proposition~\ref{p:loss-perf} (for perfect answers) or Proposition~\ref{p:loss-imp} (for imperfect answers).

If a source is capable of perfect answers (for instance, in the simple linear model) solution of problem (\ref{eq:minLV}) reduces to finding the efficient frontier: if $L^*(G)$ is the expression describing the efficient frontier (abstracting from its true discrete structure) and $Y_s$ is the capacity of the information source, then the minimum in (\ref{eq:minLV}) is equal to $L^*(Y_s)$ and is achieved by the question $\bC$ lying on the efficient frontier such that $G(\Omega,\bC,P)=Y_s$.

If a source cannot provide perfect answers (likely a more realistic scenario), one would need to consider questions with difficulty exceeding the source capacity ($G(\Omega,\bC,P)>Y_s$) in order to minimize the expected loss. The search for an optimal question in this case becomes somewhat more complicated as the error structure for the source's answers needs to be taken into account. If answers are assumed, for instance, to be quasi-perfect, optimal question(s) can be readily found approximately provided the efficient frontier is already known. An illustration is provided in the next section.

\begingroup
\squeezetable
\begin{table*}
\begin{tabular}{|c|c|l|}\hline
Pseudoenergy & Loss & Comments \\\hline\hline
$G(\Omega,\bC_f(P),P)$ & $L(g_P,P)$ & exhaustive question difficulty/total initial loss (EVPI) \\\hline
$G(\Omega,\bC,P)$ & $S(x_P^*,P_{\bC})=B(g_{\bC,P},P)$ & question difficulty/loss reduction due to perfect answer \\\hline
$Y(\Omega,\bC,P,V(\bC))$ & $S(x_P^*,P_{V(\bC)})$ & answer depth/loss reduction due to that answer\\\hline
$G(\Omega,\bC,P_{V(\bC)})$ & $B(g_{\bC,P},P_{V(\bC)})$ & residual difficulty/``price'' of answer imperfection\\\hline
$G(\Omega,\bC_f(P)_{\bC},P)$ & $L(g_{\bC,P},P)$ & residual pseudoenergy/loss given perfect answer to $\bC$\\\hline
$G(\Omega,\bC_f(P),P_{V(\bC)})$ & $\sum_{k=1}^m v_k L(g_{P^k},P^k)$ & residual pseudoenergy/loss given an imperfect answer to $\bC$ \\\hline
\end{tabular}
\caption{\label{t:energy-loss}Correspondence between pseudoenergy and loss related quantities.}
\end{table*}
\endgroup

\begin{figure}
\includegraphics[scale=0.8]{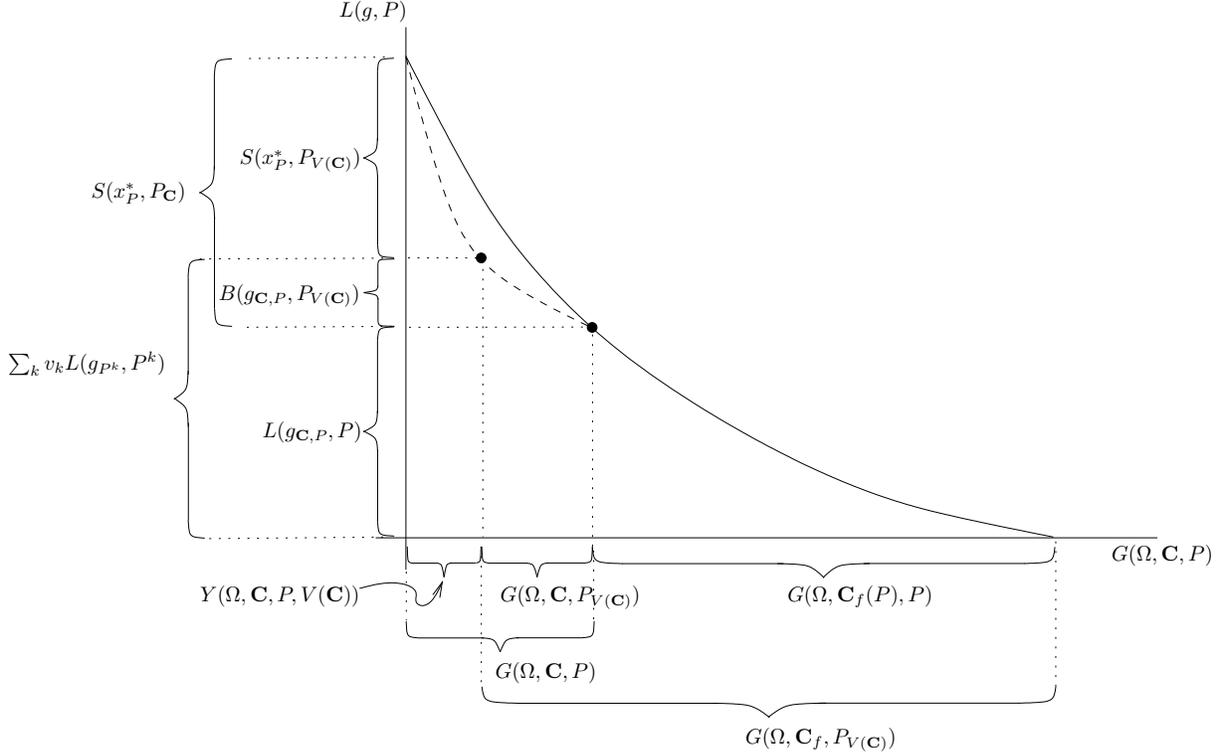}
\caption{\label{f:energylosscurve}The efficient frontier and correspondence between pseudoenergy and objective function (loss) quantities. A Pareto-optimal map $g\in \mO$ on the
efficient frontier is shown.}
\end{figure}

The correspondence between pseudoenergy and loss quantities shown in Table~1 can be illustrated by comparing decompositions of the exhaustive question difficulty $G(\Omega,\bC_f(P),P)$ (expression (\ref{eq:energy-decomp})) and  the EVPI $L(g_P,P)$ (expression (\ref{eq:loss-decomp})) on the other hand. It is also shown in Fig.~\ref{f:energylosscurve}.
\begin{equation}
\begin{aligned}
&\hphantom{Y(\Omega,\bC,P,V(\bC))+}\overbrace{\hphantom{G(\Omega,\bC,P_{V(\bC)})+
G(\Omega,\bC_f(P)_{\bC},P)}}^{G(\Omega,\bC_f(P),P_{V(\bC)})}  \\[-6pt]
&\underbrace{Y(\Omega,\bC,P,V(\bC))+G(\Omega,\bC,P_{V(\bC)})}_{G(\Omega,\bC,P)}+G(\Omega,\bC_f(P)_{\bC},P)
=G(\Omega,\bC_f(P),P)
\end{aligned}
\label{eq:energy-decomp}
\end{equation}

\begin{equation}
\begin{aligned}
&\hphantom{S(x_P^*,P_{V(\bC)})+}\overbrace{\hphantom{B(g_{\bC,P},P_{V(\bC)})+L(g_{\bC,P},P)}}^{\sum_{k=1}^m v_k L(g_{P^k},P^k)}\\[-6pt]
&\underbrace{S(x_P^*,P_{V(\bC)})+B(g_{\bC,P},P_{V(\bC)})}_{S(x_P^*,P_{\bC})=B(g_{\bC,P},P)}+
L(g_{\bC,P},P)=L(g_P,P)
\end{aligned}
\label{eq:loss-decomp}
\end{equation}

\section{\label{s:example}Example}
Suppose a company has to decide on the order quantity $x$ of a certain product and is required to satisfy an uncertain demand $\o$. The cost of ordering is $c>0$ per unit of product. If the demand is larger than the ordered quantity, the shortage has to be covered by back ordering at a higher cost $b>c$. If the demand turns out to be lower than the ordered quantity, the extra units are held in storage at unit cost of $h>0$. Thus the total cost has the form
\begin{equation}
f(\o,x)=cx+b[x-\o]_+ + h[\o-x]_+,
\label{eq:total-cost}
\end{equation}
where $[y]_+=\max\{y,0\}$ for any real $y$. We assume that both $x$ and $\o$ are continuous variables, for convenience. It is well-known that if the measure on the parameter space $\Omega$ is described by a cdf $F(\cdot)$ then the optimal solution of the problem
\begin{equation}
\mbox{min}_x \bE_P f(\o, x),
\label{eq:pr-inv}
\end{equation}
is given by $x_P^*=F^{-1}\left(\frac{b-c}{b+h} \right)$.

Let us assume that the probability  measure $P$ is uniform on $\Omega=[0,a]$. Then, clearly,  $x_P^*=a\frac{b-c}{b+h}$ (and therefore $g_P(\o)=a\frac{b-c}{b+h}$ for all $\o\in \Omega$). Consider partitions of $\Omega$ such that $P(C_j)=w_j$, $j=1,\dotsc, r$ and all sets $C_j$ are connected. Just like in the previous example, we can assume, without loss of generality that $C_j=[a\tilde w_j, a(\tilde w_j+w_j)]$, where $\tilde w_j=\sum_{l=1}^{j-1} w_l$ if $j>1$ and $\tilde w_1=0$.

\begin{figure}
\includegraphics[scale=0.6]{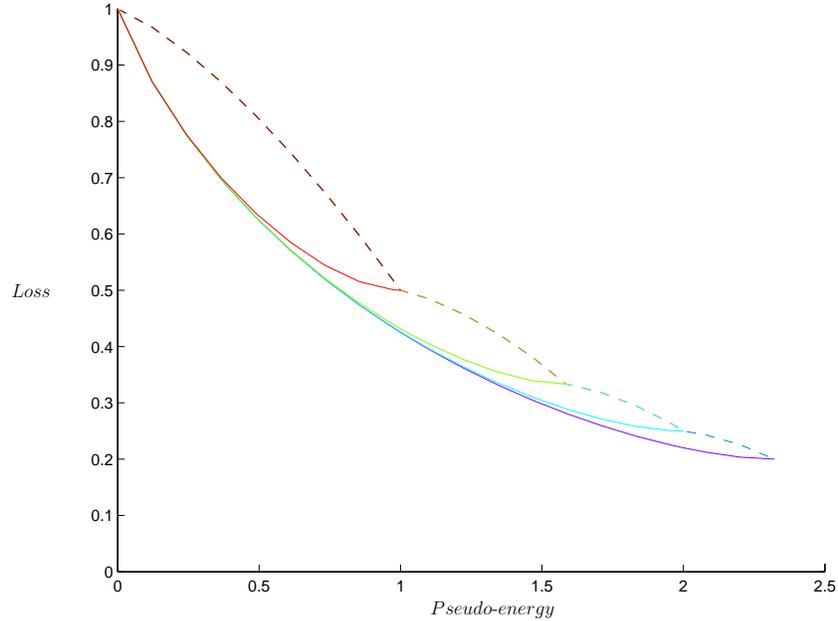}
\caption{\label{f:effront-inv}Efficient frontier for the inventory example: constant pseudotemperature case (dotted line) and linear increasing  pseudotemperature case (solid line).}
\end{figure}

It is straightforward to show that the EVPI of this problem is
\begin{equation*}
L(g_P,P)=\frac{a}{2}\cdot \frac{(b-c)(c+h)}{b+h}.
\end{equation*}
and, for the partition $\bC=\{C_1,\dotsc, C_r\}$, $x^*_{P_{C_j}}=a\left(\tilde w_j + w_j\frac{b-c}{b+h}\right)$, and
\begin{align*}
L(g_{\bC,P},P)&=L(g_{\bC,P},P_{\bC})=\sum_{j=1}^r P(C_j) L(g_{\bC,P}, P_{C_j})\\
 &= \sum_{j=1}^r w_j \frac{aw_j}{2}\cdot \frac{(b-c)(c+h)}{b+h}\\
&= \frac{a}{2}\cdot \frac{(b-c)(c+h)}{b+h} \sum_{j=1}^r w_j^2 = \left(\sum_{j=1}^r w_j^2\right) L(g_P,P)
\end{align*}
 Fig.~\ref{f:effront-inv} shows the efficient frontier for the case of constant pseudotemperature function which leads to $u(C_j)=1$ for $j=1,\dotsc, r$ and for the case of linear increasing pseudotemperature function $u(\o)=\frac{2}{a}\o$ which leads to $u(C_j)=2\tilde w_j +w_j$, $j=1,\dotsc, r$.

Let us now consider quasi-perfect answers $V_{\alpha}(\bC)$ to question $\bC$ with partitions $\bC$ as described before. Consider the case $r=2$ only, for simplicity.
Then $C_1=[0,w_1a]$ and $C_2=[w_1a, a]$. The optimal solutions to (\ref{eq:pr-inv}) with the original measure $P$ replaced with $P^k$ can be shown to be
\begin{equation}
x^*_{P^1}=
\begin{cases}
\frac{w_1a}{1-\alpha w_2}\cdot \frac{b-c}{b+h} & \text{if $\alpha < \frac{1}{w_2}\cdot \frac{c+h}{b+h}$} \\
\frac{a}{\alpha}\left(\alpha-\frac{c+h}{b+h} \right) & \text{if $\alpha \ge \frac{1}{w_2}\cdot \frac{c+h}{b+h}$},
\end{cases}
\end{equation}
and
\begin{equation}
x^*_{P^2}=
\begin{cases}
a\left(1-\frac{w_2}{1-\alpha w_1}\cdot \frac{c+h}{b+h} \right) & \text{if $\alpha < \frac{1}{w_1}\cdot \frac{b-c}{b+h}$} \\
\frac{a}{\alpha}\cdot \frac{b-c}{b+h} & \text{if $\alpha \ge \frac{1}{w_1}\cdot \frac{b-c}{b+h}$}.
\end{cases}
\end{equation}
The suboptimalities $S(x^*_P, P^k)$ for $k=1,2$ can then be calculated. The resulting expressions are too lengthy (and not very illuminating) to be given here. The resulting loss can be found as
\begin{equation}
B(g_{\bC,P},P_{V(\bC)})+L(g_{\bC,P},P)=L(g_P,P)-S(x_P^*,P_{V(\bC)}),
\label{eq:inv-loss-imp}
\end{equation}
and the pseudoenergy content of answer $V_{\alpha}(\bC)$  is simply $Y(\Omega,\bC,P,V_{\alpha}(\bC))$ given by (\ref{eq:Y-qp}). Let us set, for definiteness, $c=1$, $b=1.5$, $h=0.1$ and $a=100$. Then the EVPI of the original problem is $L(g_P,P)=17.19$. Let us also consider two information sources, described by the modified linear model, with equal capacity of $Y_s=0.2$ (in the average unit pseudotemperature calibration) and same value of parameter $b=0.8$. The first source is characterized by a constant pseudotemperature function $u(\o)\equiv 1$ and the second has linear increasing pseudotemperature $u(\o)=\frac{2}{a}\cdot \o$. The second source can be said to have relatively more ``knowledge'' about lower values of possible demand.

\begin{figure}
\includegraphics[scale=0.3]{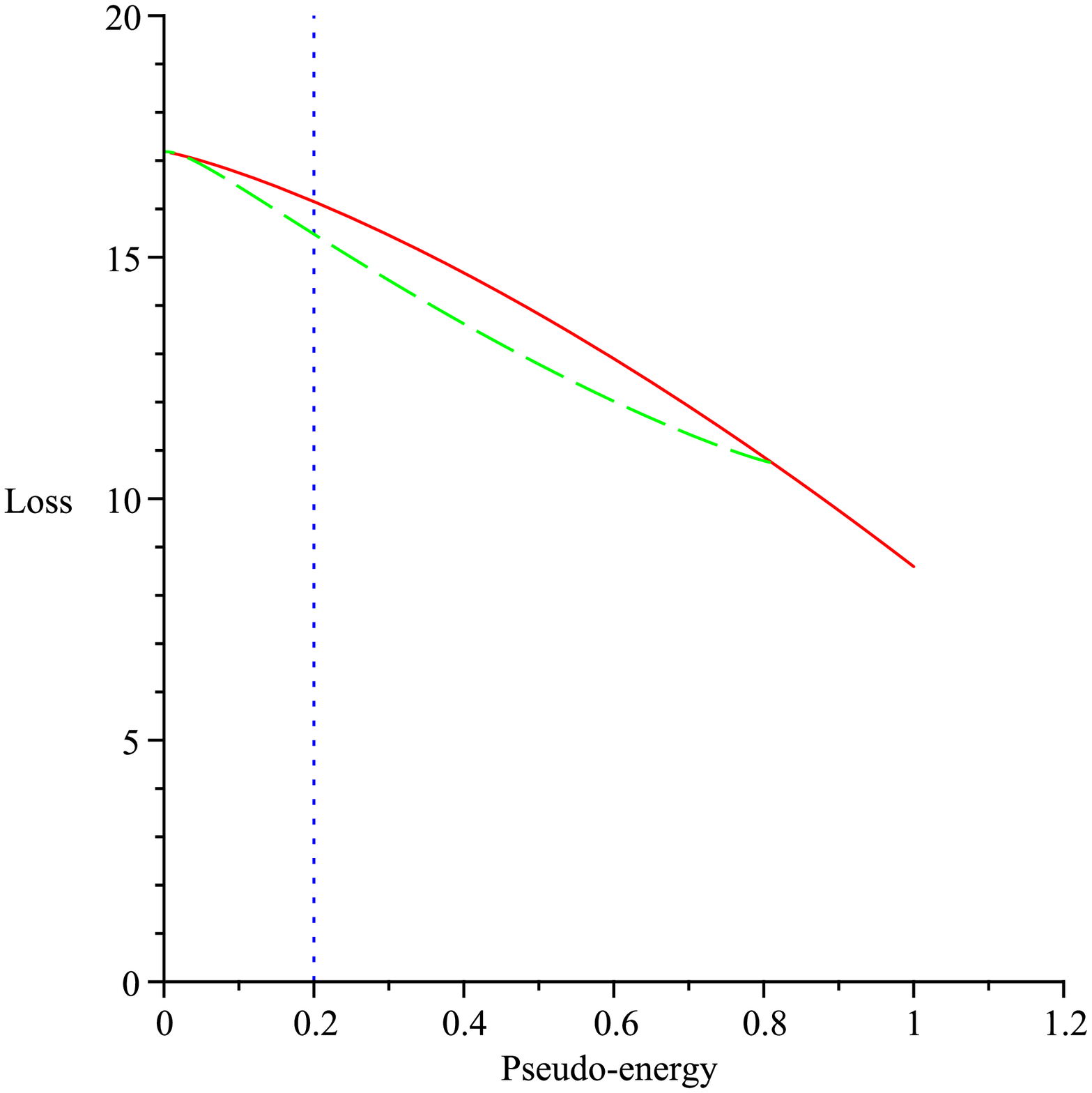}
\includegraphics[scale=0.3]{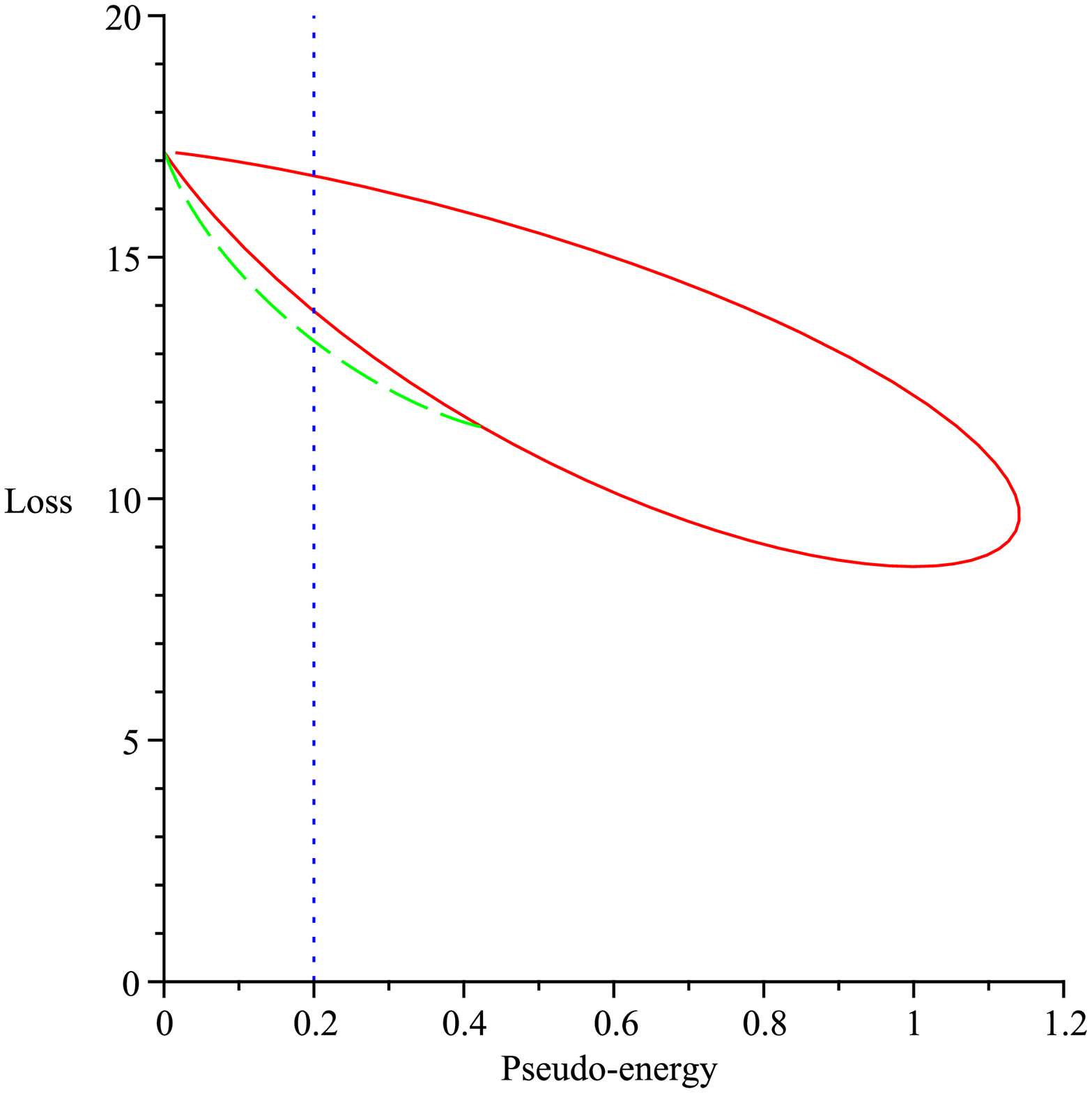}
\caption{\label{f:inventory-imp}Loss vs. pseudoenergy (for $r=2$ questions only) for a source with constant pseudotemperature (left) and a source with linear increasing pseudotemperature (right). On both plots, the solid line is obtained by varying the parameter $w_1$ from 0 to 1. The dashed line is obtained by fixing a value of $w_1$ and varying $\alpha$ from 0 to 1. The value of $w_1$ (characterizing the optimal question) is chosen so that the point of intersection of the dashed line and the vertical dotted line (source capacity) has the lowest possible value of the vertical coordinate. The latter is equal to the minimum expected loss $\mL(V(\bC))$.}
\end{figure}

We are interested in finding, for each source, an $r=2$ question $\bC=\{C_1,C_2\}$ an answer to which would help the agent minimize the expected loss. This can easily be done numerically, for example, by graphing the loss (\ref{eq:inv-loss-imp}) against the answer depth $Y(\Omega,\bC,P,V_{\alpha}(\bC))$, for different questions $\bC$ (in this case, uniquely characterized by a single parameter $w_1$). It turns out (see Fig.~\ref{f:inventory-imp} for an illustration) that the minimum loss at $Y(\Omega,\bC,P,V_{\alpha}(\bC))=Y_s=0.2$ is achieved for $w_1=0.25$ for the first source and $w_1=0.21$ for the second source. The minimum loss itself turns out to be equal to $\mL(V(\bC))=15.48$ for the first source and $\mL(V(\bC))=13.27$ for the second source, representing, respectively, $10\%$ and $23\%$ loss reduction from the original EVPI of 17.19. Clearly, the reason the second source is able to help the agent significantly more is that the latter is capable of utilizing the particular ``expertise'' of the second source by asking a question that is easy for the source and thus can be answered relatively well (with error probability $\alpha=0.21$). On the other hand, the first source answers its ``best'' question with  error probability of $\alpha=0.56$ which results -- expectedly -- in a lower loss reduction. Note that the difficulty of the optimal question is equal to 0.80 for the first source and 0.41 for the second source, while the depth of the respective answer is equal to 0.2 (the source's capacity) in both cases. Note also that, in the modified linear model, a source can provide an answer of depth equal to capacity $Y_s$ whenever the question difficulty exceeds the value $Y_s/b$, i.e. the question has to be sufficiently difficult for the source so that the latter can provide an answer of maximum depth.

\section{\label{s:conclusion}Conclusion}
Despite the role information plays in science and engineering, the fundamental theory of information itself is still largely limited to just the middle link of the {\it full information chain} that generally includes information acquisition, transmission and usage stages (links). The theory of the middle link -- the classical Information Theory -- describes information transmission and can be concisely characterized as a theory of information {\it quantity}. If a description of the end links of the information chain is desired, a theory of information {\it accuracy} and {\it relevance} is required.

This article is devoted to development of the basic framework of a theory of the information usage link.
Since the two end links are closely logically connected, they have to be treated together, and the results of \cite{part1,part2} on the basics of the information acquisition link are used here to arrive at the formulation of the {\it optimal information acquisition problem} which, in its elementary form, searches for an optimal question to the given information source needed to maximize the solution quality (understood as loss reduction) for the given (optimization) problem. Such a question can be thought of as a way of achieving an optimal ``alignment'' between the information source (the first link) and the problem (the third link), for the given state of ``information background'' -- the initial probability measure.

Solving  the optimal information acquisition problem is facilitated by consideration of the Pseudoenergy-Loss efficient frontier in the space of all possible questions. The latter consists of all questions that are the most relevant for the given problem among all that are at most as difficult for the given source. The knowledge of the efficient frontier enables the agent to (approximately) find optimal questions for a source of with a known {\it knowledge structure} (described by question difficulty functional) and {\it pseudoenergy capacity} which can be given an interpretation of the sources maximum knowledge depth. It is interesting to observe that the two end links of the information chain exhibit a notable symmetry, with pseudoenergy (accuracy) and loss (relevance) quantities coming in corresponding pairs. One can talk of a duality between the two links. This duality appears to be a manifestation of the tight interconnection between the end links.

Finally, the problem of finding the efficient frontier of questions appears to be a computationally difficult one. Fortunately, it turns out that methods based on probability metrics which were used in scenario reduction approaches to stochastic optimization can also be of use for approximate efficient frontier determination. This is the main subject of the companion paper \cite{loss-part2}.

\appendix

\section{\label{a:qas}Questions, Answers and Source Models}

\subsection{Questions and their difficulty}
A definition of questions was originally given by Cox in \cite{cox1979}. There, a questions was associated with a set of all logical assertions that answer it fully. This line was further developed in \cite{knuth05} where a distributive lattice of questions was constructed from the lattice of logical assertions, questions being associated with {\it down-sets} of subsets of elements of the latter lattice.
In the context of our discussion, a question so defined would be associated with an {\it inclusion-free}\footnote{A collection of subsets of $\Omega$ is called inclusion-free if neither member of such a collection is a subset of another.} collection of subsets of $\Omega$. Moreover, {\ real questions} of \cite{cox1979} and \cite{knuth05} are associated with inclusion-free collections of subsets of $\Omega$ that cover the whole of $\Omega$, and {\it partition questions} (which will be of primary interest to us) correspond to inclusion-free subsets that are partitions of $\Omega$, i.e. do not include overlapping subsets. Besides standard (complete) partitions, we also make use of {\it incomplete partitions}, i.e. collections of non-overlapping  subsets of $\Omega$ that do not fully cover $\Omega$. We call questions associated with them {\it incomplete questions}. Additionally, if such a partition consists of a single subset of $\Omega$, the corresponding question is called, following \cite{knuth05}, an {\it ideal question}.

A {\it difficulty functional} $G(\Omega,\bC,P)$ can be associated with any question $\bC=\{C_1,\dotsc, C_r\}$. The particular form of $G(\Omega,\bC,P)$ can be determined if some reasonable requirements are imposed. This was done in \cite{part1} where a particular system of postulates expressing  {\it linearity} and {\it isotropy} properties of the difficulty functional was proposed. The main theorem proved in \cite{part1} derives the general form of the difficulty functional that is required to satisfy such postulates.
\begin{theorem}
Let the functional $G(\Omega, \bC, P)$ where $\bC=\{C_1,\dotsc, C_r\}$ satisfy Postulates 1 through 6 (see \cite{part1}). Then it has the form
$$G(\Omega, \bC, P)=\frac{\sum_{j=1}^r u(C_j)P(C_j)\log \frac{1}{P(C_j)}}{\sum_{j=1}^r P(C_j)},$$
where $u(C_j)=\frac{\int_{C_j}u(\o)\,dP(\o)}{P(C_j)}$ and $u$: $\Omega\rightarrow \bR$  is an integrable nonnegative function on the parameter space~$\Omega$.
\label{th:G}
\end{theorem}

In particular, the difficulty of the given question $\bC$ depends on, besides the initial probability measure $P$, the function $u(\cdot)$ defined on the parameter space $\Omega$. This function may be called the {\it pseudotemperature} using parallels with thermodynamics. The question difficulty then can be interpreted as the amount of {\it pseudoenergy} associated with question~$\bC$.

If $\tilde\bC$ is an arbitrary refinement\footnote{A refinement of a partition of $\Omega$ is another partition such that every member of it is a subset of some member of the original partition.} of $\bC$ then the difficulty of the more detailed question $\tilde\bC$ can be decomposed as (\cite{part1})
\begin{equation}
G(\Omega,\tilde\bC,P)= G(\Omega,\bC,P)+G(\Omega,\tilde\bC_{\bC},P),
\label{eq:seq_c}
\end{equation}
where the expected {\it residual difficulty} of $\tilde \bC$ given a perfect answer to $\bC$ is defined as
\begin{equation}
G(\Omega,\tilde\bC_{\bC},P)=\sum_{C\in \bC}P(C)G(C, \tilde\bC_C, P_C).
\label{eq:chain}
\end{equation}

\subsection{Answers and their depth}
Given a question $\bC$ on $\Omega$, an answer to $\bC$ was defined in \cite{part2} to be a message $V(\bC)$ taking values in the set $\{s_1, \dotsc, s_m\}$ such that the reception of the value $s_k$ modifies (updates) the initial measure $P$ on $\Omega$ to the measure $P^k$ such that $P^k_{C_j}=P_{C_j}$ (whenever conditional measures are defined) for $k=1,\dotsc, m$ and $j=1,\dotsc, r\}$. The latter condition ensures that the answer $V(\bC)$ is indeed an answer to the question $\bC$ (and no more).

It follows from the above definition that, for $V(\bC)$ to be an answer to a complete question $\bC$, it is necessary and sufficient for the updated measures $P^k$, $k=1,\dotsc, m$, to take the form
\begin{equation}
P^k=\sum_{j=1}^r p_{kj}P_{C_j},
\label{eq:Pk-mc}
\end{equation}
where $p_{kj}$, $k=1,\dotsc, m$, $j=1,\dotsc, r$ are nonnegative coefficients such that $\sum_{j=1}^r p_{kj}=1$ for $k=1,\dotsc, m$. The expression (\ref{eq:Pk-mc}) is modified somewhat \cite{part2} for incomplete questions. The probability of an answer $V(\bC)$ taking value $s_k$ is denoted by $v_k$. It is assumed that the updated measures $P^k$, $k=1,\dotsc, m$, are consistent with the original measure $P$
in the sense that
\begin{equation}
\sum_{k=1}^m v_kP^k = P.
\label{eq:vkPk=P}
\end{equation}
Informally speaking, the condition (\ref{eq:vkPk=P}) means that the original measure $P$ is a ``valid'' one which is only ``refined'' by the information source's answers.

The {\it answer depth functional} $Y(\Omega,\bC,P, V(\bC))$ for the answer $V(\bC)$ to question $\bC$ measures the amount of {\it pseudoenergy} that is conveyed by $V(\bC)$ in response to question $\bC$. The general form of $Y(\Omega,\bC,P, V(\bC))$ can be established if certain requirements it has to satisfy are imposed. This was done in \cite{part2} where postulates expressing {\it linearity} and {\it isotropy} properties were formulated. Under these conditions, the following result was obtained.
\begin{theorem}
The answer depth functional $Y(\Omega, \bC, P, V(\bC))$ has the form
$$Y(\Omega, \bC, P, V(\bC))=\sum_{k=1}^m \Pr(V(\bC)=s_k)\frac{\sum_{j=1}^r u(C_j)P^k(C_j)\log \frac{P^k(C_j)}{P(C_j)}}{\sum_{j=1}^r P^k(C_j)}, $$
where $P^k$ is the measure on $\Omega$ updated by the reception of $V(\bC)=s_k$ and
$u(C_j)=\frac{1}{P(C_j)}\int_{C_j} u(\o) dP(\o)$  and the function $u$: $\Omega\rightarrow \bR$ is the same function that is used in the question difficulty functional $G(\Omega, \bC,P)$.
\label{th:Y}
\end{theorem}

It can be shown (see \cite{part2} for details) that if $V(\bC)$ is any answer to the question $\bC$ then $Y(\Omega, \bC, P, V(\bC))\le G(\Omega, \bC, P)$  with equality if and only if the answer $V(\bC)$ is {\it perfect}, i.e. $P^j=P_{C_j}$ for $j=1,\dotsc, r$. The difficulty of question $\bC$ can be written as
\begin{equation}
G(\Omega,\bC,P)=Y(\Omega,\bC,P,V(\bC))+G(\Omega,\bC,P_{V(\bC)}),
\label{eq:G=Y+G}
\end{equation}
where
\begin{equation}
G(\Omega,\bC,P_{V(\bC)})=\sum_{k=1}^m v_k \sum_{j=1}^r u(C_j)P^k(C_j)\log \frac{1}{P^k(C_j)}
\label{eq:res-diff}
\end{equation}
can be termed the {\it residual difficulty} of $\bC$ given the answer $V(\bC)$. Clearly, $G(\Omega,\bC,P_{V(\bC)})\ge 0$ with the inequality being tight for a perfect answer $V^*(\bC)$.
The residual difficulty $G(\Omega,\bC,P_{V(\bC)})$ can be expressed via coefficients $p_{kj}$ that describe the answer $V(\bC)$:
\begin{equation}
G(\Omega,\bC,P_{V(\bC)})=\sum_{k=1}^m \sum_{j=1}^r v_k p_{kj}u(C_j)\log \frac{1}{p_{kj}}
\label{eq:Gres-pkj}
\end{equation}

It turns out to be convenient to consider the class of imperfect answers for which the degree of imperfection is described by a single error probability $\alpha$ -- the {\it quasi-perfect} answers \citep{part2}. For a quasi-perfect answer $V_{\alpha}(\bC)$ to a (complete) question $\bC=\{C_1,\dotsc, C_r\}$, the coefficients $p_{kj}$ have the form
\begin{equation}
 p_{kj}=(1-\alpha)\delta_{k,j}+\alpha P(C_j),
 \label{eq:pkj-alpha}
 \end{equation}
 for $k=1,\dotsc, r$ and $j=1,\dotsc, r$, and the updated measure $P^k$ is simply
 \begin{equation}
 P^k = \alpha P + (1-\alpha) P_{C_k}.
 \label{eq:Pk-alpha}
 \end{equation}
 for $k=1,\dotsc, r$.
 Clearly, for $\alpha=0$ a quasi-perfect answer to $\bC$ becomes a perfect one. It can be shown (see \cite{part2}) that the answer depth functional for a quasi-perfect answer $V_{\alpha}(\bC)$ to question $\bC$ can be written as
 \begin{equation}
 \begin{split}
 Y(\Omega,\bC,P,V_{\alpha}(\bC))&=\sum_{k=1}^r u(C_k)P(C_k)(1-\alpha +\alpha P(C_k))\log \frac{1-\alpha +\alpha P(C_k)}{P(C_k)}\\
 &+ \alpha \log \alpha \sum_{k=1}^r u(C_k)P(C_k)(1-P(C_k)),
 \end{split}
 \label{eq:Y-qp}
 \end{equation}
 which is easily seen to reduce to $G(\Omega,\bC,P)$ for $\alpha=0$ and vanish for $\alpha=1$.

\subsection{Information source models}
The pseudotemperature function $u(\cdot)$ on the parameter space $\Omega$ characterizes (under the linear isotropic model considered here) the source specific relative difficulty of questions ``located'' in various regions of $\Omega$. An information source model relates the value of answer depth to the difficulty of the corresponding question. Formally speaking, the existence of information source models is based on the following hypothesis \cite{part2}.

{\bf Hypothesis S1}. For the given information source and any question $\bC$, the answer depth is a function of the question difficulty:
$$Y(\Omega,\bC,P,V(\bC))=h(G(\Omega,\bC,P)),$$
where $h$ : $\bR_+\rightarrow \bR_+$ is a function of a single argument.

The simplest information source model considered in \cite{part2} is the simple capacity model given by
\begin{equation}
h(x)=\begin{cases} x & \mbox{if}\; x\le Y_s \\
                   Y_s & \mbox{if}\; x>Y_s.
\end{cases}
\label{eq:Y(G)-cap-simple}
\end{equation}
which is fully characterized by the single parameter $Y_s$ which has the meaning of the information source capacity.

The most apparent drawback of model (\ref{eq:Y(G)-cap-simple}) is that it predicts that the source would provide a perfect answer to any question whose difficulty does not exceed the source capacity. The linear modified capacity model described by
\begin{equation}
h(x)=\begin{cases} bx & \mbox{if}\; x\le \frac{Y_s}{b} \\
                   Y_s & \mbox{if}\; x>\frac{Y_s}{b}
\end{cases}
\label{eq:Y(G)-cap-mod}
\end{equation}
removes this drawback at the expense of one extra parameter $b\le 1$ that has to be estimated. Several slightly more complicated models were proposed in \cite{part2}.

The values of model parameters as well as pseudotemperature functions for information sources can be estimated from the observed sources' performance on some set of sample questions. Optimization based formulations for such estimation were also proposed in \cite{part2}.

It is easy to see that multiplying the pseudotemperature function $u(\cdot)$ by a constant has the effect of multiplying both the question difficulty and the answer depth by the same constant and is equivalent to a choice of units of pseudoenergy. It turns out to be convenient to use two different conventions in this regard.
\begin{itemize}
\item The convention in which $\int_{\Omega} u(\o)\, d\o =1$. Here the units of pseudoenergy are chosen in such a way that, for constant $u(\o)$, the pseudoenergy coincides with entropy making it convenient to make use of the standard intuition about entropy and information.

\item The convention in which each source has unit capacity ($Y_s=1$). This choice of units of pseudoenergy makes it convenient to compare the ``depth of knowledge'' of different information sources to each other by directly comparing their respective pseudotemperature values at the same points of the parameter space.
\end{itemize}

\section{\label{a:proofs}Proofs}

\subsection{Proof of Lemma~\ref{l:loss1}}
\begin{align*}
L(g,P)&= \int_{\Omega} (f(\omega, g(\omega))-f(\omega, x_{\omega}^*)) P(d\omega)\\
      &= \sum_{j=1}^r \int_{C_j} (f(\omega, g(\omega))-f(\omega, x_{\omega}^*)) P(d\omega) \\
      &=  \sum_{j=1}^r P(C_j) \int_{C_j} \frac{1}{P(C_j)}(f(\omega, g(\omega))-f(\omega, x_{\omega}^*))P(d\omega)\\
      &=\sum_{j=1}^r P(C_j) \int_{C_j} (f(\omega, g(\omega))-f(\omega, x_{\omega}^*))P_{C_j}(d\omega)\\
      &\overset{(a)}{=}  \sum_{j=1}^rP(C_j) L(g,P_{C_j}) \overset{(b)}{=} L(g,P_{\bC}),
\end{align*}
where (a) follows directly from the definition of the expected loss for the measure $P_{C_j}$ and (b) follows from the definition (\ref{eq:L(bC)}) of $L(g,P_{\bC})$.

\subsection{Proof of Lemma~\ref{l:L=S+L}}
We have
\begin{align*}
L(g_P,P) &= \int_{\Omega} (f(\o, x_P^*)-f(\o, x_{\o}^*))P(d\o) \\
         &= \int_{\Omega} (f(\o, x_P^*)-f(\o, x_{\o}^*)+f(\o, g_{\bC,P}(\o))-f(\o, g_{\bC,P}(\o)))P(d\o)\\
         &= \int_{\Omega} (f(\o, x_P^*)-f(\o, g_{\bC,P}(\o)))P(d\o)\\
         &+ \int_{\Omega} (f(\o, g_{\bC,P}(\o))-f(\o, x_{\o}^*))P(d\o) \\
         &= \sum_{j=1}^r P(C_j) \int_{C_j} \frac{1}{P(C_j)}\left( f(\o, x_P^*)-f(\o, g_{\bC,P}(\o)) \right)P(d\o)\\
         &+ \sum_{j=1}^r P(C_j) \int_{C_j} \frac{1}{P(C_j)}\left( f(\o, g_{\bC,P}(\o))-f(\o, x_{\o}^*) \right) P(d\o) \\
         &\overset{(a)}{=} \sum_{j=1}^r P(C_j) \int_{C_j} \left( f(\o, x_P^*)-f(\o, x_{P_{C_j}}^*) \right) P_{C_j}(d\o)\\
         &+ \sum_{j=1}^r P(C_j) \int_{C_j} \left( f(\o, g_{\bC,P}(\o))-f(\o, x_{\o}^*) \right) P_{C_j}(d\o) \\
         &\overset{(b)}{=} \sum_{j=1}^r P(C_j) S(x_P^*,P_{C_j})+\sum_{j=1}^r P(C_j) L(g_{\bC,P},P_{C_j})\\
         &\overset{(c)}{=}S(x_P^*,P_{\bC})+L(g_{\bC,P},P_{\bC}) \overset{(d)}{=}S(x_P^*,P_{\bC})+L(g_{\bC,P},P) ,
\end{align*}
where (a) follows from the definition of the conditional measure $P_{C_j}$, (b) follows from the definitions of $S(x_P^*,P_{C_j})$ and $L(g,P_{C_j})$, (c) follows from the notational convention  (\ref{eq:f(bC)}) for functionals of measures, and (d) follows from Lemma~\ref{l:loss1}.

\subsection{Proof of Proposition~\ref{p:loss-perf}}
For the given value $s_k$ of the answer, $P^j=P_{C_j}$, $j=1,\dotsc, r$. Therefore the agent can achieve the smallest possible loss by choosing the solution $x^*_{P_{C_j}}$. The resulting conditional loss will be
\begin{equation}
\mL(s_j)=\int_{C_j} (f(\o, x^*_{P_{C_j}})-f(\o,x^*(\o)))\, dP_{C_j}(\o).
\label{eq:mLsj}
\end{equation}
Taking the expectation of (\ref{eq:mLsj}) over possible values of the answer $V^*(\bC)$ we obtain
\begin{align*}
\mL(V^*(\bC))&\overset{(a)}{=}\sum_{j=1}^r P(C_j)\mL(s_j)= \sum_{j=1}^r P(C_j) \int_{C_j} (f(\o, x^*_{P_{C_j}})-f(\o,x^*_{\o}))\, dP_{C_j}(\o)\\
&\overset{(b)}{=} \sum_{j=1}^r P(C_j) \int_{C_j} (f(\o, g_{\bC,P}(\o))-f(\o,x^*_{\o}))\, dP_{C_j}(\o)\\
&= \sum_{j=1}^r P(C_j)L(g_{\bC,P},P_{C_j})\overset{(c)}{=}L(g_{\bC,P},P_{\bC})
\overset{(d)}{=}L(g_{\bC,P},P),
\end{align*}
where (a) follows from that for a perfect answer consistent with the original measure,  $\Pr(V^*(\bC)=s_j)=P(C_j)$, (b) follows from that the map $g_{\bC,P}$ is subset-optimal, (c) follows from the definition (\ref{eq:L(bC)}), and (d) follows from Lemma~\ref{l:loss1}.

\subsection{Proof of Lemma~\ref{l:S=S+G}}
\begin{align*}
S(x_P^*,P_{\bC})&=\sum_{j=1}^r P(C_j)S(x_P^*, P_{C_j})\\
 &= \sum_{j=1}^r P(C_j) \int_{C_j} (f(\o, x_P^*) - f(\o, g_{\bC,P}(\o)))P_{C_j}(d\o)\\
&\overset{(a)}{=} \sum_{j=1}^r \sum_{k=1}^r p_{kj}v_k \int_{C_j} (f(\o, x_P^*) - f(\o, g_{\bC,P}(\o)))P_{C_j}(d\o)\\
&\overset{(b)}{=} \sum_{j=1}^r \sum_{k=1}^r p_{kj}v_k \int_{\Omega} (f(\o, x_P^*) - f(\o, g_{\bC,P}(\o)))P_{C_j}(d\o)\\
&= \sum_{k=1}^r v_k \int_{\Omega} \sum_{j=1}^r p_{kj} (f(\o, x_P^*) - f(\o, g_{\bC,P}(\o)))P_{C_j}(d\o)\\
&\overset{(c)}{=} \sum_{k=1}^r v_k \int_{\Omega} (f(\o, x_P^*) - f(\o, g_{\bC,P}(\o)))P^k(d\o)\\
&= \sum_{k=1}^r v_k \int_{\Omega} (f(\o, x_P^*) - f(\o, g_{\bC,P}(\o))+ f(\o, x_{P^k}^*)-f(\o, x_{P^k}^*))P^k(d\o)\\
&= \sum_{k=1}^r v_k \int_{\Omega} (f(\o, x_P^*) - f(\o, x_{P^k}^*))P^k(d\o)\\
&+ \sum_{k=1}^r v_k \int_{\Omega} (f(\o, x_{P^k}^*) - f(\o, g_{\bC,P}(\o)))P^k(d\o)\\
&\overset{(d)}{=} \sum_{k=1}^r v_k S(x_P^*,P^k)+\sum_{k=1}^r v_k B(g_{\bC,P},P^k)\\
&\overset{(e)}{=}S(x_P^*,P_{V(\bC)})+B(g_{\bC,P},P_{V(\bC)}),
\end{align*}
where (a) follows from the consistency condition (\ref{eq:vkPk=P}), (b) follows from the fact that measure $P_{C_j}$ vanishes outside of $C_j$, (c) follows from the form (\ref{eq:Pk-mc}) of the updated measures $P^k$, (d) follows from the definitions (\ref{eq:subopt}) and (\ref{eq:gain}) of suboptimality and gain, and (e) follows from the definitions (\ref{eq:S(V)}) and (\ref{eq:B(V)}).

\subsection{Proof of Proposition~\ref{p:loss-imp}}
The value $s_k$ of answer $V(\bC)$ implies that the measure on $\Omega$ is equal to $P^k$. Therefore the the agent can achieve minimum loss by using the stochastic optimal solution $x^*_{P^k}$. The resulting minimum loss will be
\begin{equation}
\mL(s_k)=L(g_{P^k},P^k),
\label{eq:mLs_k}
\end{equation}
where $g_{P^k}$ is the all-to-one map $g_{P^k}(\o)=x^*_{P^k}$ for all $\o\in \Omega$.

The minimum expected loss $\mL(V(\bC))$ can be obtained by substituting (\ref{eq:mLs_k}) into (\ref{eq:mLV}):
\begin{equation}
\mL(V(\bC))=\sum_{k=1}^m {v_k}L(g_{P^k},P^k).
\label{eq:mLV(bC)}
\end{equation}

On the other hand, we can decompose the EVPI $L(g_P,P)$ as follows.
\begin{align}
&L(g_P,P) = \int_{\Omega} (f(\o,x_P^*)-f(\o,x_{\o}^*))P(d\o)\nonumber\\
&= \sum_{k=1}^m v_k \int_{\Omega} (f(\o,x_P^*)-f(\o,x_{\o}^*)) P^k(d\o)\nonumber \\
&= \sum_{k=1}^m v_k \int_{\Omega} (f(\o,x_P^*)-f(\o,x_{\o}^*)+f(\o,x_{P^k}^*)-f(\o,x_{P^k}^*)) P^k(d\o)\nonumber\\
&=  \sum_{k=1}^m v_k \int_{\Omega} (f(\o,x_P^*)-f(\o,x_{P^k}^*)) P^k(d\o)\nonumber\\
&+ \sum_{k=1}^m v_k \int_{\Omega} (f(\o,x_{P^k}^*)-f(\o,x_{\o}^*)) P^k(d\o)\nonumber\\
&= \sum_{k=1}^m v_k S(x_P^*,P^k) + \sum_{k=1}^m v_k L(g_{P^k},P^k)
= S(x_P^*,P_{V(\bC)}) + \sum_{k=1}^m v_k L(g_{P^k},P^k)
\label{eq:L=S+L'}
\end{align}

Comparing (\ref{eq:mLV(bC)}) with (\ref{eq:L=S+L'}) we can obtain
\begin{equation}
\mL(V(\bC))=L(g_P,P)-S(x_P^*,P_{V(\bC)}).
\label{eq:mLV=L-S}
\end{equation}

Finally, using the decomposition of EVPI of Corollary~\ref{c:L=S+B+L} in (\ref{eq:mLV=L-S}) yields
\begin{equation*}
\mL(V(\bC))=B(g_{\bC,P},P_{V(\bC)})+L(g_{\bC,P},P).
\end{equation*}

\section{\label{a:maps}Examples of maps}
Let $\Omega$ be the interval $[0,a]$ and let $X$ be the real line $\bR$. Let the integrand $f(\o,x)$ have the following form: $f(\o,x)=(x-\o)^2$ and let the original measure $P$ be the uniform continuous distribution on $[0,a]$.

It is obvious that the optimal solution for the given realization $\o$ is simply $x^*_{\o}=\o$. The stochastic optimal map is $g_P(\o)=\frac{a}{2}\in X$ for all $\o\in \Omega$. Therefore the EVPI of the problem (\ref{eq:gen_stoch}) is
\begin{equation*}
L(g_P,P)=\frac{1}{a}\int_0^a \left((x_P^*-\o)^2 - (x_{\o}^*-\o)^2 \right) d\o=\frac{1}{a}\int_0^a \left(\frac{a}{2}-\o \right)^2 d\o=\frac{a^2}{12}.
\end{equation*}

\begin{figure}
\includegraphics[scale=0.6]{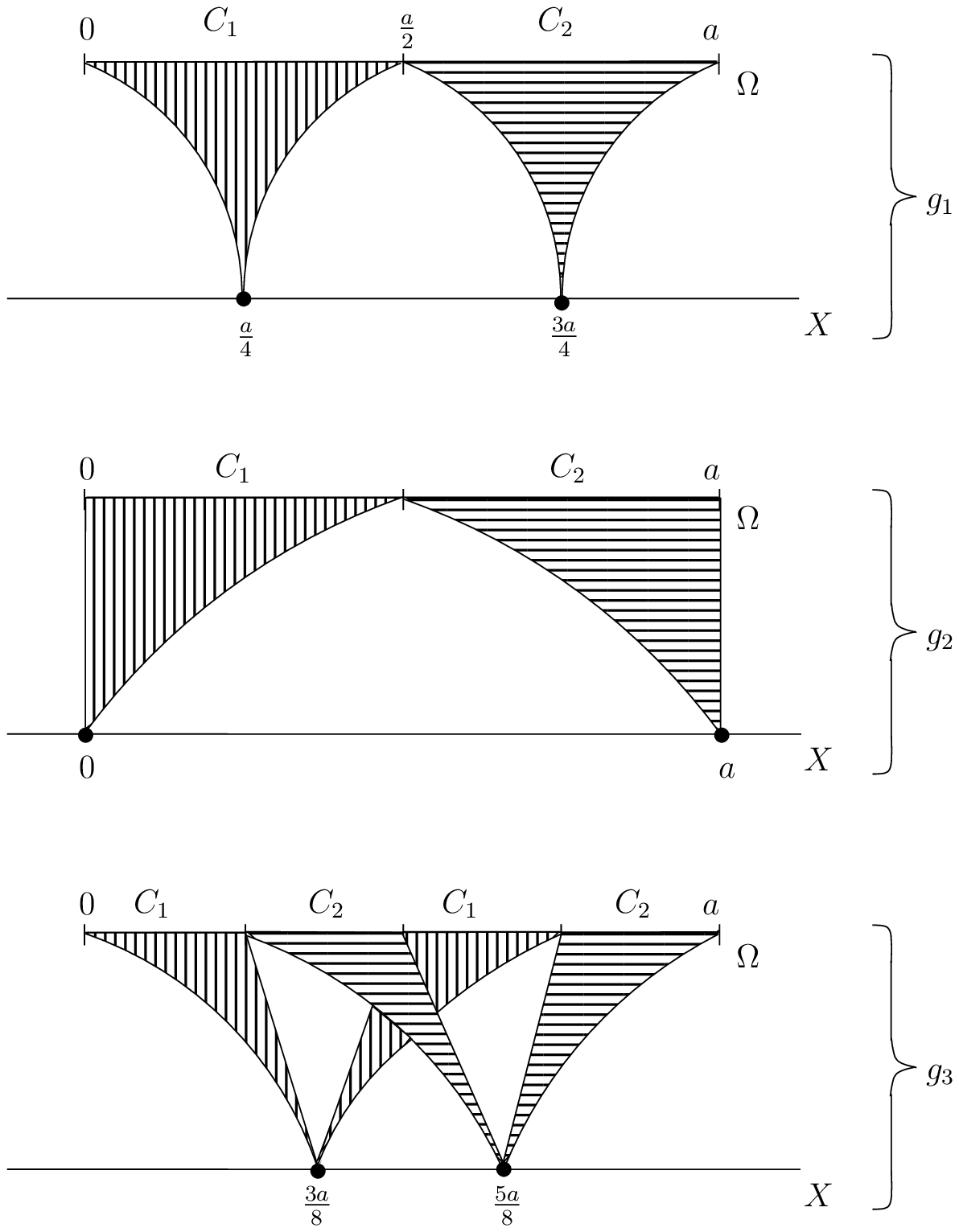}
\caption{\label{f:3maps}Maps $g_1$, $g_2$ and $g_3$. The partitions for $g_1$ and $g_2$ consist of connected sets only. Each element of the partition for $g_3$
consists of two connected sets.}
\end{figure}

\begin{figure}
\includegraphics[scale=0.6]{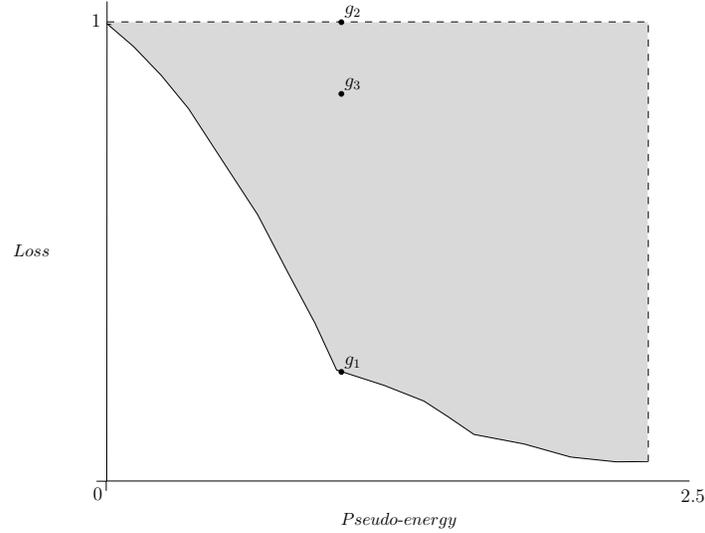}
\caption{\label{f:3mapsHL}Maps $g_1$, $g_2$ and $g_3$ on $(G,L)$ coordinate plane. All possible maps for this problem lie in the shaded region,
at or above the efficient frontier.}
\end{figure}

Let $\bC=\left\{\left[0,\frac{a}{2}\right), \left[\frac{a}{2},a \right] \right\}$ and $\bC'= \left\{\left[0,\frac{a}{4}\right)\cup \left[\frac{a}{2}, \frac{3a}{4}\right) ,\left[\frac{a}{4}, \frac{a}{2}\right)\cup \left[\frac{3a}{4},a\right]\right\}$ be two $r=2$ partitions of $\Omega$.
Let us consider several different $r=2$ maps $g\in \mG$  (see Fig.~\ref{f:3maps} for an illustration).
\begin{itemize}
\item $g_1=\left(\bC,\left\{\frac{a}{4}, \frac{3a}{4}\right\}\right)=g_{\bC,P}$. The measures $P_{C_1}$ and $P_{C_2}$ are uniform on $C_1$ and $C_2$ respectively. We have $x_{P_{C_1}}^*=\frac{a}{4}$ and $x_{P_{C_2}}^*=\frac{3a}{4}$. Thus $g_1\in \mC$. Note that in this case $g_1\in \mO$ as well as it lies on the efficient frontier in $(G,L)$ coordinate plane (see Fig.~\ref{f:3mapsHL} for an illustration).

\item $g_2=\left(\bC,\{0,a\}\right)$. For this map, the partition is the same as that for $g_1$, but the image set is different. This map is therefore not subset-optimal: $g_2\notin \mC$.

\item $g_3=\left(\bC',\left\{ \frac{3a}{8}, \frac{5a}{8} \right\}\right)= g_{\bC',P}$. For this map's partition both subsets $C'_1$ and $C'_2$ consist of two connected components. It is easy to check that $x_{P_{C_1}}^*=\frac{3a}{8}$ and $x_{P_{C_2}}^*=\frac{5a}{8}$ and thus $g_3\in \mC$.
\end{itemize}

The loss for these three maps can be found as follows.
For $g_1$, $$L(g_1,P)=\frac{1}{2}\cdot \frac{2}{a}\int_{0}^{a/2} \left(\frac{a}{4}-\o \right)^2 d\o + \frac{1}{2}\cdot \frac{2}{a}\int_{a/2}^{a} \left(\frac{3a}{4}-\o \right)^2 d\o = \frac{a^2}{48},$$
for $g_2$,
$$L(g_2,P)=\frac{1}{2}\cdot \frac{2}{a}\int_{0}^{a/2} \left(0-\o \right)^2 d\o + \frac{1}{2}\cdot\frac{2}{a} \int_{a/2}^{a} \left(1-\o \right)^2 d\o = \frac{a^2}{12},$$
and for $g_3$,
\begin{align*}
L(g_3,P) &= \frac{1}{2}\cdot \frac{2}{a}\left( \int_{0}^{a/4} \left(\frac{3a}{8}-\o \right)^2 d\o + \int_{a/2}^{3a/4} \left(\frac{3a}{8}-\o \right)^2 d\o  \right)\\
&+ \frac{1}{2}\cdot\frac{2}{a}\left( \int_{a/4}^{a/2} \left(\frac{5a}{8}-\o \right)^2 d\o + \int_{3a/4}^{a} \left(\frac{5a}{8}-\o \right)^2 d\o \right)  = \frac{13a^2}{192}.
\end{align*}

Fig.~\ref{f:3mapsHL} shows the efficient frontier and maps $g_1$, $g_2$ and $g_3$ in $(G,L)$ coordinate plane. We see that $g_1\in \mO$ lies on the efficient frontier while $g_2$ and $g_3$ are located above it.

Since $g_1, g_3\in \mC$ we have (as Lemma~\ref{l:L=S+L} states) $S(x_P^*,P_{\bC})=\frac{a^2}{12}-\frac{a^2}{48} = \frac{a^2}{16}$ for $g_1$ and $S(x_P^*,P_{\bC'})=\frac{a^2}{12}-\frac{13a^2}{192} = \frac{a^2}{64}$ for $g_3$. For $g_2$, the suboptimality is the same as that for $g_1$. Note that, since $g_2\notin \mC$, $S(x_P^*,P_{\bC})+L(g_3,P)=\frac{7a^2}{48}\ne L(g_P,P)$.

For this one-dimensional example it turns out to be straightforward to find maps on the efficient frontier. Indeed, it is obvious that partitions for such maps have to consist of connected sets only. It is also clear that the order in which subsets $C_j$ appear on the interval $[0,a]$ does not matter because the integrand in (\ref{eq:gen_stoch}) $f(\o,x)$ depends on $|\o-x|$ only. So, for the fixed value of $r$, any map $g\in \mC$ that can lie on the efficient frontier can be uniquely characterized by the subset measures $w_j=P(C_j)$, $j=1,\dotsc, r$. Given the values $w_j$, the expected loss of the corresponding map can be written as
\begin{equation*}
L(g,P)=\sum_{j=1}^r w_j\frac{(w_ja)^2}{12}=\frac{a^2}{12}\sum_{j=1}^r w_j^3.
\end{equation*}
In order to find the optimal values of $w_j$ yielding the smallest loss for the question difficulty $G(\Omega,\bC,P)$ not exceeding $h$ the following optimization problem needs to be solved.
\begin{equation}
\begin{aligned}
& \text{minimize}
& & \sum_{j=1}^r w_j^3  \\
& \text{subject to}
& & -\sum_{j=1}^r u(C_j)w_j\log w_j\le h \\
& & &\sum_{j=1}^r w_j =1 \\
& & & w_j\ge 0, \quad j=1,\dotsc,r,
\end{aligned}
\label{eq:opt_wj}
\end{equation}
where $u(C_j)$ is the pseudotemperature of subset $C_j$ and $h$ is a nonnegative parameter. Since the function $-\sum_{j=1}^r u(C_j)w_j\log w_j$ is concave, (\ref{eq:opt_wj}) is a global optimization problem. However it can easily be solved to optimality for moderate values of the partition size $r$. We consider two cases: constant pseudotemperature function $u(\o)\equiv 1$ and linear pseudotemperature $u(\o)=\frac{2}{a}\o$. We can assume that $C_j=[a\tilde w_j, a(\tilde w_j+w_j)]$. In the former case, $u(C_j)=1$, $j=1,\dotsc, r$ and in the latter case,
\begin{equation}
u(C_j)=2\tilde w_j +w_j,
\label{eq:uni-weights}
\end{equation}
where $\tilde w_j=\sum_{l=1}^{j-1} w_l$ if $j>1$ and $\tilde w_1=0$.

\begin{figure}
\includegraphics[scale=0.6]{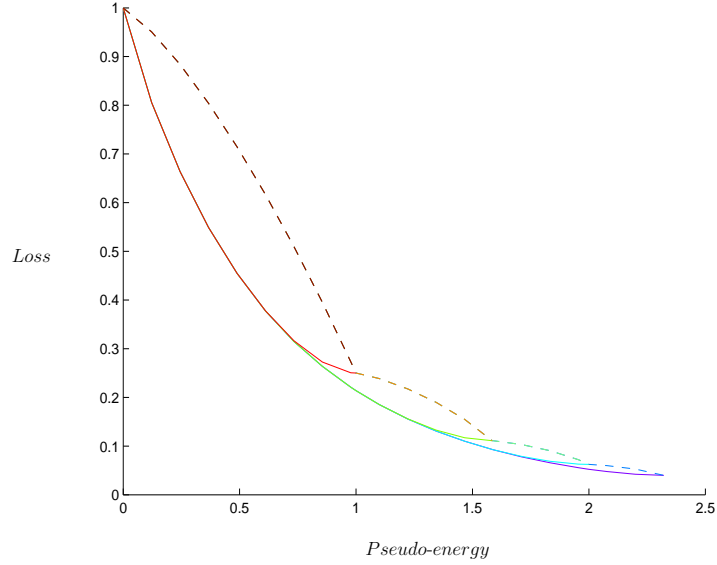}
\caption{\label{f:eff-front_ex}Efficient frontier for the toy example: constant pseudotemperature case (dotted line) and linear pseudotemperature case (solid line).}
\end{figure}

The resulting efficient frontier is shown in Fig.~\ref{f:eff-front_ex}.

Let us now consider imperfect answers to questions $\bC$ in the same example. For simplicity, we set $r=2$ for questions and assume the pseudotemperature to be constant on $\Omega$. We also assume all answers to be quasi-perfect so that the updated measures $P^k$, $k=1,2$ have the form (\ref{eq:Pk-alpha}).

The stochastic optimal solutions $x^*_{P^k}$ for measures $P^k$ can be found as
\begin{equation*}
x^*_{P^k} = \arg\min_x \int_{\Omega} f(\o,x) P^k(d\o).
\end{equation*}
We have
\begin{align*}
x^*_{P^1}&= \arg\min_x \left( \frac{1-\alpha(1-w_1)}{w_1a} \int_{0}^{w_1a} (x-\o)^2 d\o +
\frac{\alpha}{a}\int_{w_1a}^{a} (x-\o)^2 d\o \right)\\
&= \frac{1}{2}(w_1a+\alpha(1-w_1)a)=\frac{1}{2}a(w_1+\alpha w_2),
\end{align*}
and, analogously,
$$x^*_{P^2}=\frac{1}{2}a(w_2+\alpha w_1).$$
We can now find the suboptimalities:
\begin{align*}
S(x_P^*,P^1)&=\int_{\Omega}\left(f(\o, x_P^*)-f(\o, x_1^*(\alpha)) \right) P_1^{(\alpha)}(d\o)\\
&= \frac{a^2}{12}\left((3-6w_1+3w_1^2)(1+\alpha^2)+\alpha(-6+12w_1-6w_1^2) \right),
\end{align*}
and, analogously,
\begin{equation*}
S(x_P^*,P^2)=\frac{a^2}{12}\left((3-6w_2+3w_2^2)(1+\alpha^2)+\alpha(-6+12w_2-6w_2^2) \right).
\end{equation*}
The suboptimality $S(x_P^*,P_{V(\bC)})$ is then
\begin{align*}
S(x_P^*,P_{V(\bC)})&= w_1S(x_P^*,P_1^{(\alpha)})+w_2S(x_P^*,P_2^{(\alpha)})\\
&= \frac{a^2}{12} (1-w_1^3-w_2^3)(1-\alpha)^2.
\end{align*}

The new value of the expected loss is
\begin{equation}
L(g_P,P)-S(x_P^*,P_{V(\bC)})=\frac{a^2}{12}-\frac{a^2}{12}(1-w_1^3-w_2^3)(1-\alpha)^2
\label{eq:loss(alpha)}
\end{equation}

Note that for $\alpha=0$ we recover the expression $L(g_{\bC,P},P)=\frac{a^2}{12}(w_1^3+w_2^3)$ for a perfect answer and for $\alpha=1$ the new value of the loss is simply $L(g_P,P)=\frac{a^2}{12}$ since $\alpha=1$ describes the case in which the answer $V(\bC)$ carries no new information and the updated measure is simply $P$.

\begin{figure}
\includegraphics[scale=0.4]{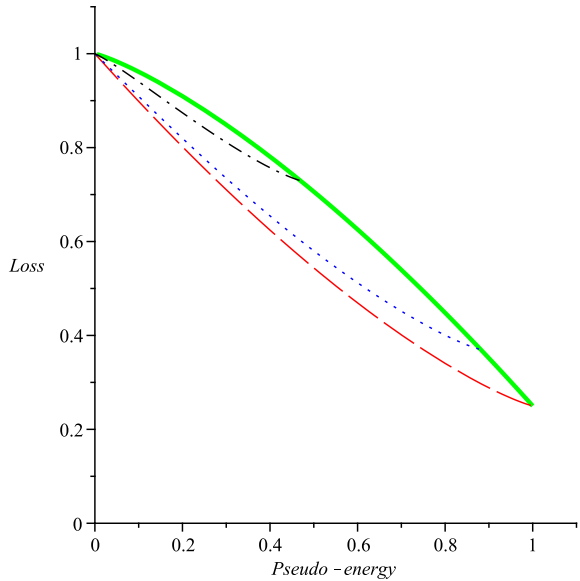}
\caption{\label{f:alpha_curves}Dependence of the expected loss on the added information for $r=2$ partitions. The solid curve corresponds to the error-free message case with $w_1$ varying from 0 to 0.5. The dashed line shows the $w_1=w_2=0.5$ case with $\alpha$ varying from 1 to 0 (from left to right on the figure). The dotted line is the same for $w_1=1-w_2=0.7$ case, and the dash-dotted line is for $w_1=1-w_2=0.9$ case.}
\end{figure}

Fig.~\ref{f:alpha_curves} shows the dependence of the expected loss (\ref{eq:loss(alpha)}) on answer depth with the error parameter $\alpha$ ranging from 0 to 1 for several values of subset measures $w_1$ and $w_2$ for the $r=2$ case. The part of the efficient frontier that can be achieved for $r=2$ is also shown (solid bold line). It is interesting to observe that, for the same amount of pseudoenergy, lower values of the expected loss can be achieved with imperfect answers to more difficult questions.

%%%%%%%%%%%%%%%%%%%%%%%%%%%%%%%%%%%%%%%%%%%%%%%%%%%%%%%%%

%\bibliography{E:/work/IT-OPT/thesis_bib}

%merlin.mbs apsrev4-1.bst 2010-07-25 4.21a (PWD, AO, DPC) hacked
%Control: key (0)
%Control: author (8) initials jnrlst
%Control: editor formatted (1) identically to author
%Control: production of article title (-1) disabled
%Control: page (0) single
%Control: year (1) truncated
%Control: production of eprint (0) enabled
%

\end{document}